\documentstyle[eqsecnum,aps,prb,multicol,graphicx]{revtex}
\begin{document}
\draft
\title{
Neutron scattering study of 
dipolar spin ice 
Ho$_2$Sn$_2$O$_7$:
Frustrated pyrochlore magnet 
}
\author{
Hiroaki Kadowaki
}
\address{
Department of Physics, Tokyo Metropolitan University, 
Hachioji-shi, Tokyo 192-0397, Japan
}
\author{
Yoshinobu Ishii
}
\address{
Advanced Science Research Center, 
Japan Atomic Energy Research Institute, 
Tokai, Ibaraki 319-1195, Japan
}
\author{
Kazuyuki Matsuhira\cite{byline} and Yukio Hinatsu
}
\address{
Division of Chemistry, Graduate School of Science, 
Hokkaido University, Sapporo, Japan
}
\date{received: May 21, 2001}
\maketitle
\begin{abstract}
By means of neutron scattering techniques 
we have investigated 
the frustrated pyrochlore magnet Ho$_2$Sn$_2$O$_7$, 
which was found to show a ferromagnetic 
spin-ice behavior below $T \simeq$ 1.4 K 
by susceptibility measurements. 
High-resolution powder-neutron-diffraction 
shows no detectable disorder of the lattice, 
which implies appearance of a random magnetic 
state solely by frustrated geometry, i.e., 
the corner sharing tetrahedra. 
Magnetic inelastic-scattering spectra show that Ho 
magnetic moments behave as an Ising spin system 
at low temperatures, 
and that 
the spin fluctuation has static character. 
The system remains in a short-range ordered state 
down to $T =$ 0.4 K. 
By analyzing the wave-number dependence of the magnetic scattering 
using a mean field theory, 
it is clarified that the Ising spins 
interact via the dipolar interaction.
Therefore we conclude that Ho$_2$Sn$_2$O$_7$ belongs to 
the dipolar-spin-ice family. 
Slow spin dynamics is exhibited as 
thermal hysteresis and time dependence of 
the magnetic scattering.
\end{abstract}


\begin{multicols}{2}
\section{
Introduction
}
\label{intro}
\begin{figure}
\begin{center}
\includegraphics[width=9.0cm,clip]{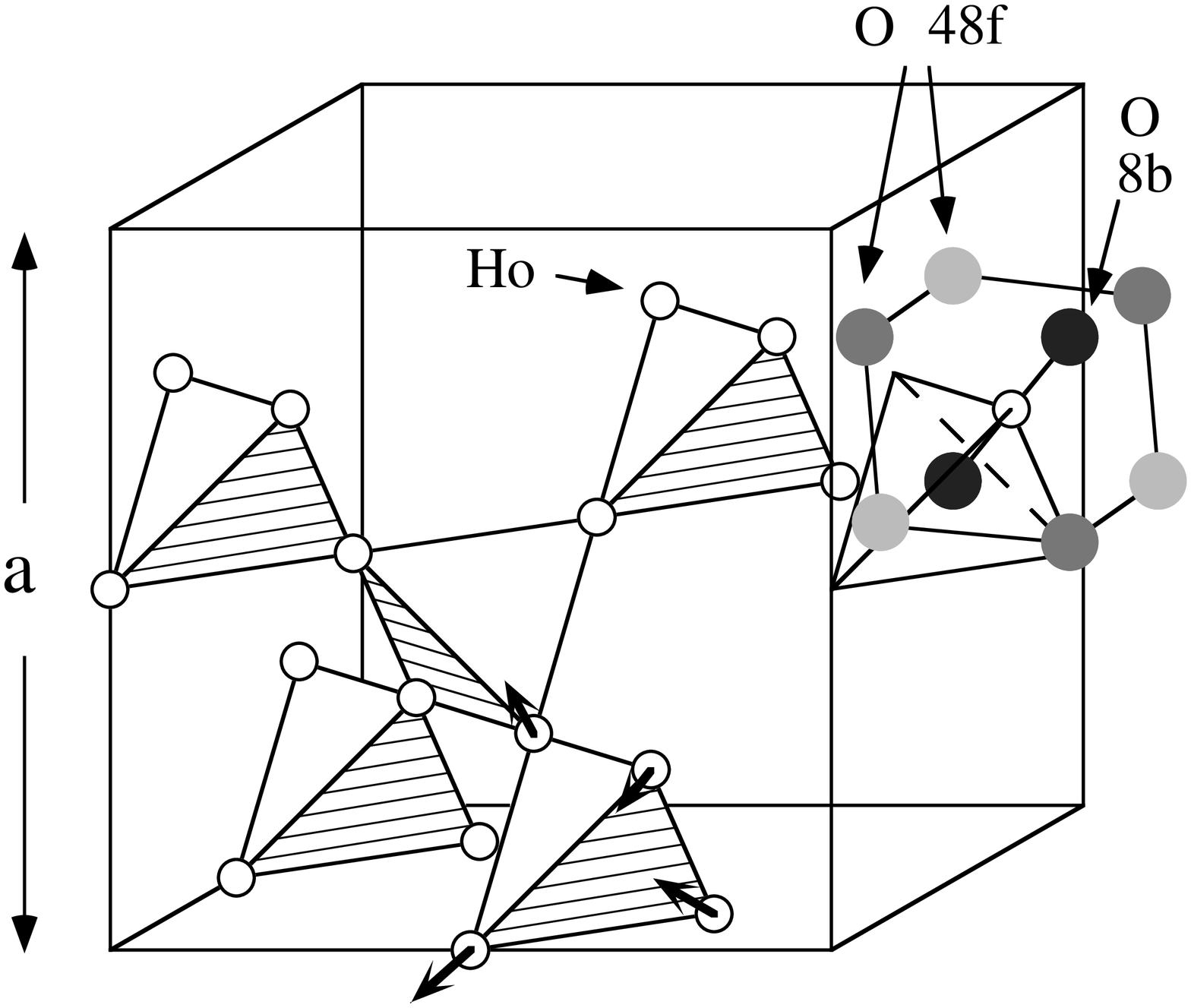}
\end{center}
\caption{
Network of corner sharing tetrahedra of Ho 
is shown in a unit cell of the pyrochlore 
Ho$_2$Sn$_2$O$_7$.
Oxygen atoms giving rise to local trigonal symmetry 
around a Ho atom, and 
``two-in-two-out'' spin configuration on 
a tetrahedron are also illustrated.
}
\label{crystal.HoSnO}
\end{figure}

Frustrated spin systems have been investigated mostly in 
antiferromagnets on geometrically 
frustrated lattices such as triangular, kagome, FCC, pyrochlore 
lattices. 
Recently an intriguing frustrated system was found 
in an Ising ferromagnet on the pyrochlore lattice 
by Harris {\it et al.}, 
\cite{Harris97,Bramwell98}
 where magnetic ions
form a lattice of corner sharing tetrahedra illustrated in 
Fig.~\ref{crystal.HoSnO}. 
When spins interact via a ferromagnetic nearest-neighbor 
exchange-interaction and 
have strong local Ising-anisotropies which force each spin to 
point into or out-of the center of each tetrahedron, 
spin configurations of the ground-state 
on one tetrahedron are sixfold 
degenerate ``two-in and two-out'' structures 
(see Fig.~\ref{crystal.HoSnO}). 
By extending this ``two-in-two-out'' structure to 
the entire pyrochlore lattice, 
it was shown that ground state degeneracy is
macroscopic, 
which leads to the residual entropy 
\cite{anderson,Ramirez99} 
at $T$ = 0. 
Since these spin configurations can be mapped to 
proton configurations in the cubic ice, \cite{anderson}
the system is 
called as the spin ice model. \cite{Harris97,Bramwell98}

Recent discovery of spin-ice model systems, 
Ho$_2$Ti$_2$O$_7$ 
\cite{Harris97,blote,Siddharthan,Rosenkranz,matsuhira,Bramwell.cm,kanada.nd}
 and Dy$_2$Ti$_2$O$_7$, \cite{Ramirez99,blote} 
which belong to a series of pyrochlore oxides 
showing frustrated properties, 
has renewed interest of ice models. 
Experimental and theoretical work of 
these compounds 
has shown that the spin system 
freezes into a certain state 
below a temperature of the order $T$ = 1 K 
\cite{Harris97,matsuhira}
which has only short range order 
\cite{Harris97,Bramwell98,Bramwell.cm}
with macroscopic number of degeneracy, i.e., zero-point entropy. 
\cite{anderson,Ramirez99,blote,Hertog}

Although the strong Ising-like anisotropies of 
Ho and Dy moments are obvious, 
the spin-spin interactions have certain ambiguity, 
because these large moments systems 
have significant contribution from dipolar interaction 
\cite{Ramirez99,Siddharthan} in addition to 
the nearest-neighbor exchange-interaction.
The long-range nature of the dipolar 
interaction complicates the real spin-ice systems, 
and another theoretical problem of a dipolar spin-ice model
\cite{Hertog,gingras.cm} 
has been addressed. 
The addition of the dipolar interaction 
removes the ground-state degeneracy 
and a long-range ordered state becomes the ground state. 
However experiments and Monte Carlo simulations 
do not show this magnetic order. 
This contradiction is thought to be 
solved by slow spin dynamics in the macroscopic number 
of low energy states 
belonging to the ``two-in-two-out'' manifold of the spin 
ice model, 
which prevents the dipolar spin ice from reaching to 
thermal equilibrium states.

Recently, other pyrochlore compounds 
Ho$_2$Sn$_2$O$_7$ and Dy$_2$Sn$_2$O$_7$ 
were shown to belong to the spin ice families
by AC susceptibility measurements. 
\cite{matsuhira} 
It was shown that Ho$_2$Sn$_2$O$_7$ 
exhibits the slow dynamics below 
a temperature scale of $T_f$ $\simeq$ 1.4 K, 
and that 
it does not have a magnetic transition to 
a long-range ordered phase down to 
$T$ = 0.15 K.

In this work, 
in order to clarify the spin ice behavior 
of Ho$_2$Sn$_2$O$_7$ on a microscopic basis, 
we have performed neutron scattering experiments 
on a powder sample.
By measuring magnetic diffraction pattern down to 
$T$ = 0.4 K, 
formation of magnetic short-range order and 
freezing effects were investigated. 
We analyzed the diffraction pattern 
using a mean field approximation 
of wave-number dependent susceptibilities, 
and evaluated inter-spin interaction constants. 
In addition, 
we measured magnetic inelastic-scattering 
spectra to elucidate energy scales of spin fluctuations.
Since a disorder of the pyrochlore lattice 
can influence the interpretation of experimental 
results in particular about the origin of the 
slow dynamics, we studied the crystal structure using 
high-resolution powder-neutron-diffraction. 

%
\section{
Experimental method
}
\label{exp.method}
A polycrystalline sample of Ho$_2$Sn$_2$O$_7$ was prepared 
by standard solid-state reaction. \cite{matsuhira}
Stoichiometric mixture of Ho$_2$O$_3$ and SnO$_2$ was heated in air at
1200 - 1400$^{\circ}$C for 3 days with intermediate regrinding
to ensure a complete reaction. 
The powder x-ray diffraction pattern of the sample indicates
that it is a single phase with the cubic pyrochlore structure.

Neutron scattering experiments on the polycrystalline sample 
were performed using triple-axis spectrometers 
C11-HER and 4G-GPTAS installed at JRR-3M JAERI (Tokai). 
Incident or final 
neutron energies were fixed at 3 or 14 meV 
using the pyrolytic-graphite (002) monochromator 
or analyzer. 
Higher-order neutrons were removed by 
the cooled Be filter or the pyrolytic-graphite filter. 
The sample was mounted in 
a liquid ${}^3$He cryostat or 
a closed cycle ${}^4{\rm He}$-gas refrigerator.
A powder-diffraction experiment was carried out 
using the high-resolution
powder-neutron-diffractometer (HRPD) 
installed at JRR-3M.
Neutrons of the wave length 
$\lambda = 1.8238$ {\AA} were selected by the Ge (331) monochromator.
%
\section{
Experimental Results
}
\label{exp.result}
%
\subsection{
Crystal structure
}
\label{cryst.str}
\begin{figure}
\begin{center}
\includegraphics[width=17.8cm,clip]{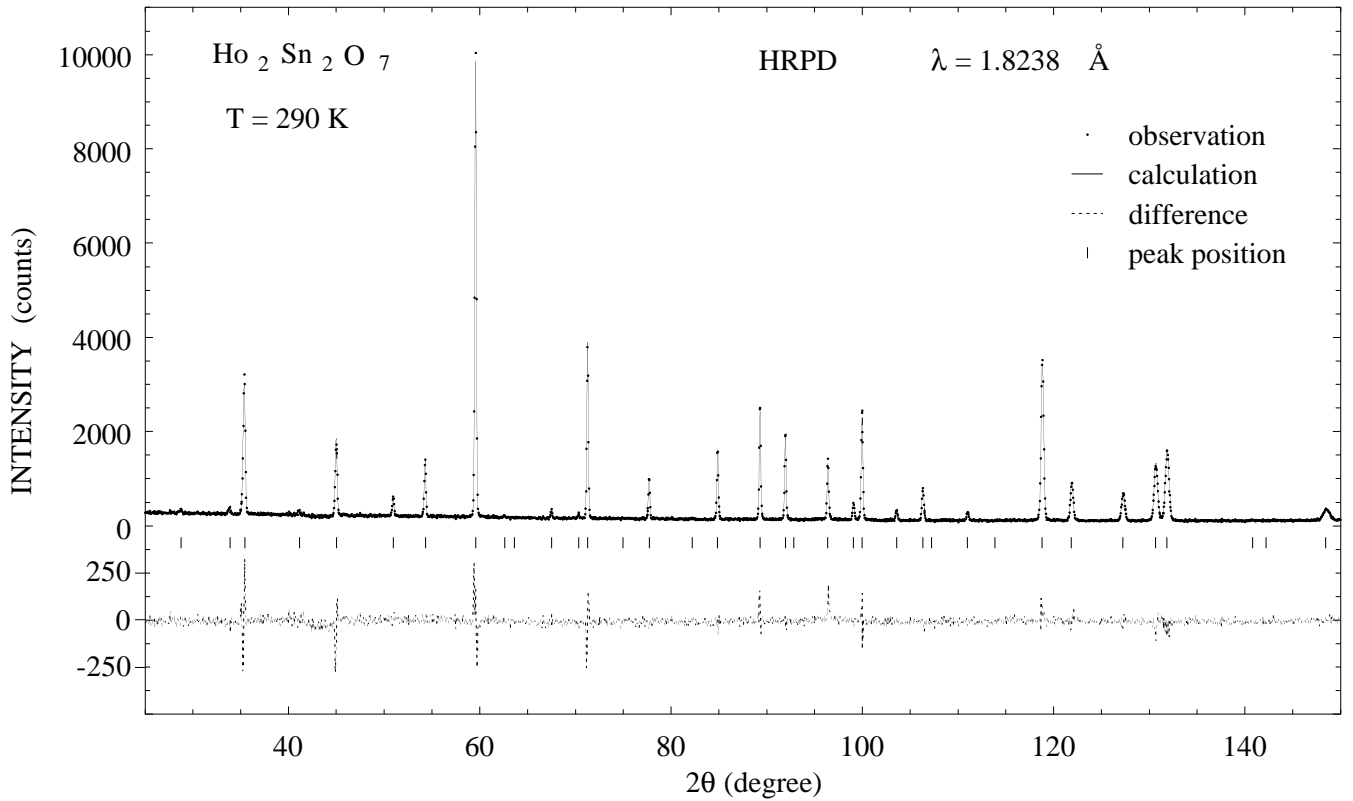}
\end{center}
\caption{
Neutron diffraction pattern of 
${\rm Ho}_2{\rm Sn}_2{\rm O}_7$ 
measured at $T=$ 290 K.
Observed and calculated patterns are denoted by 
closed circles and solid line, respectively. 
Their difference is plotted in lower part by dashed line.
Vertical bars stand for positions of Bragg reflections.
}
\label{powd.pat}
\end{figure}

The crystal structure of Ho$_2$Sn$_2$O$_7$ was studied by 
x-ray diffraction. \cite{matsuhira}
It was shown that the powder pattern is consistent with the 
fully ordered cubic pyrochlore structure. 
This structure belongs to the space group $Fd\bar{3}m$ (No. 227), 
and constituent atoms fully occupy the sites of 
16$d$(Ho), 16$c$(Sn), 48$f$(O), and 8$b$(O'). 
\cite{reimers.YMoO}
To confirm this more precisely and detect certain randomness 
in the lattice, 
we measured a powder-neutron-diffraction pattern 
at a room temperature.
The observed pattern is shown in Fig.~\ref{powd.pat}.
This was analyzed by using the Rietveld profile refinement 
program {\small RIETAN-97}. \cite{izumi}
We performed the profile fitting first 
by assuming the fully ordered pyrochlore structure. 
The resulting structure parameters are listed in 
Table \ref{tab.str}, 
and the fitted powder pattern and the difference curve
are shown in Fig.~\ref{powd.pat}. 
From this figure and 
the final $R$ factors of 
$R_{\rm wp} = 7.8$ \% ($R_{\rm e} = 6.4$ \%), 
$R_{\rm p} = 5.8$ \% and 
$R_{\rm B} = 3.5$ \%, 
we conclude that the quality of the fit is excellent. 

This refinement suggests that we do not need to introduce any 
randomness of the lattice to improve the refinement, 
or that the experimental data do not contain enough information 
to pursue small deviation from the pyrochlore structure. 
Thus we checked only one possibility which is commonly observed 
in oxides, that is, deficiency of oxygen atoms. 
A profile fitting with adjustable occupation parameters of 
48$f$(O) and 8$b$(O') sites was carried out. 
The consequent occupations were 
$n$(48$f$) = 1.00(3) and $n$(8$b$) = 0.93(7). 
This implies that determination of the oxygen deficiency 
from the present powder diffraction data is rather limited. 
We note in particular that
since the 8$b$(O') site oxygen is located at the center of the 
tetrahedron (see Fig.~\ref{crystal.HoSnO}), 
small amount of deficiency of this oxygen might 
substantially affect the Ising anisotropy of the Ho spin.
Although more precise measurements or different techniques are 
required to examine small randomness, we conclude that 
the present diffraction data are consistent with the fully ordered 
pyrochlore structure.
\begin{table}
\caption{
Refined structure parameters of 
${\rm Ho}_2{\rm Sn}_2{\rm O}_7$ at $T=$ 290 K 
using fully ordered pyrochlore structure. 
Number in parenthesis is standard deviation of the last digit.
}
\label{tab.str}
\begin{tabular}{cccccc}
  \multicolumn{6}{c}{
$Fd\bar{3}m$ (No. 227) \ \ $a=$ 10.381(1) \AA 
} \\
Atom & Site  & $x$ & $y$ & $z$ & $B$ (${\rm \AA}^2$) \\
\tableline
Ho & 16$d$ & 1/2 & 1/2  & 1/2 & 0.3(2) \\
Sn & 16$c$ & 0   & 0    & 0   & 0.3(2) \\
O  & 48$f$ & 0.3368(5) & 1/8  & 1/8 & 0.5(1) \\
O' &  8$b$ & 3/8       & 3/8  & 3/8 & 0.3(3) \\
\end{tabular}
\end{table}
\end{multicols}
%
\begin{multicols}{2}
%
%
\subsection{
Magnetic excitation
}
\label{mag.exct}
\begin{figure}
\begin{center}
\includegraphics[width=8.6cm,clip]{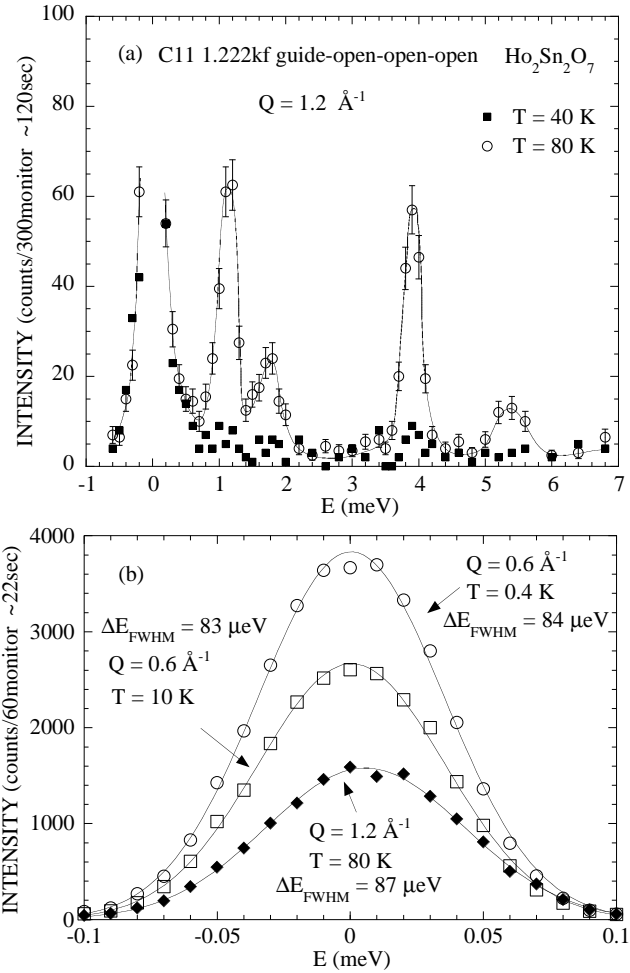}
\end{center}
\caption{
(a) Spectrum of energy scan of Ho$_2$Sn$_2$O$_7$ 
measured at $T$ = 40 and 80 K 
in range -1 $< E <$ 7 meV. 
Solid line is guide to the eye.
(b) Detail of typical elastic scattering peaks for 
0.4 $<T<$ 80 K. 
Solid lines represent fits to the Gaussian function.
Scans of (a) and (b) were carried out using the horizontally 
focusing analyzer with $E_{\rm f}$ = 3.1 meV.
}
\label{Escan.low}
\end{figure}

Magnetic excitation spectra were measured to examine 
characteristic energy scales of magnetic fluctuations 
at low temperatures and 
crystal field excitations. 
In Fig.~\ref{Escan.low}(a) 
we show excitation spectra in a low 
energy range  -1 $< E <$ 7 meV 
measured using the C11 spectrometer with a condition 
of the horizontal focusing analyzer with 
$E_{\rm f} =$ 3.1 meV. 
The excitation peaks observed at $T =$ 80 K are transitions 
between crystal-field energy levels excited thermally. 
Below $T =$ 40 K they almost vanish because of 
negligible thermal occupations, and 
magnetic signals are observed only near the elastic channel. 
A few typical energy scans near $E =$ 0 are shown 
in Fig.~\ref{Escan.low}(b).
By fitting these peaks to the Gaussian form, 
we obtained energy widths of 
$\Delta E_{\rm FWHM}$ =
83 $\pm$ 8, 84 $\pm$ 8, and 88 $\pm$ 8 $\mu$eV
as shown in the figure. 
Since these values are in agreement with 
the energy resolution $\Delta E_{\rm FWHM} =$ 85 $\mu$eV 
measured using the standard vanadium, 
we conclude that the magnetic signal near $E = $ 0 
do not show any intrinsic inelasticity 
within the present experimental 
condition. 
The upper limit of the intrinsic energy width is
$\Delta E_{\rm FWHM} <$ 30 $\mu$eV, which is 
0.3 K in the temperature scale. 
These observations are consistent with 
the strong Ising anisotropy of the spin system.

\begin{figure}
\begin{center}
\includegraphics[width=8.6cm,clip]{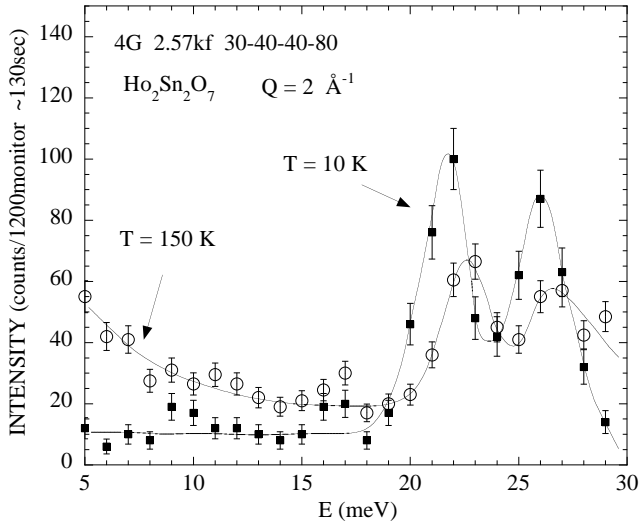}
\end{center}
\caption{
Crystal-field excitation spectrum of Ho$_2$Sn$_2$O$_7$ 
measured at 
$Q$ = 2 \AA$^{-1}$ in energy range 5 $< E <$ 30 meV.
Solid lines are guides to the eye.
}
\label{Escan.high}
\end{figure}

The crystal-field excitation spectra were measured in a wider 
energy range using the 4G spectrometer with 
a condition $E_{\rm f} =$ 14 meV. 
In Fig.~\ref{Escan.high} energy scans in a range 
5 $<E<$ 30 meV at $Q =$ 2 \AA$^{-1}$ are shown. 
One can see clearly that there are two excitation peaks at 
$E =$ 22 and 26 meV. 
Temperature variation of the spectra elucidates 
magnetic origin of these peaks. 
The positions of these crystal-field excitations 
coincide with those observed in the similar compound
Ho$_2$Ti$_2$O$_7$. \cite{Rosenkranz,Siddharthan}
For Ho$_2$Ti$_2$O$_7$ the crystal-field state has been 
thoroughly studied by neutron inelastic scattering, and the 
strong Ising character of the ground doublet is established. 
\cite{Rosenkranz}
Because of similarity between the Ti- and Sn- compounds, 
we may conclude that 
the ground doublet of Ho$_2$Sn$_2$O$_7$ 
has also the strong Ising anisotropy 
along the local $\langle 111 \rangle$ axes
and it behaves 
as an Ising model at low temperatures. 

%
\subsection{
Magnetic elastic scattering
}
\label{SRO}
\begin{figure}
\begin{center}
\includegraphics[width=8.6cm,clip]{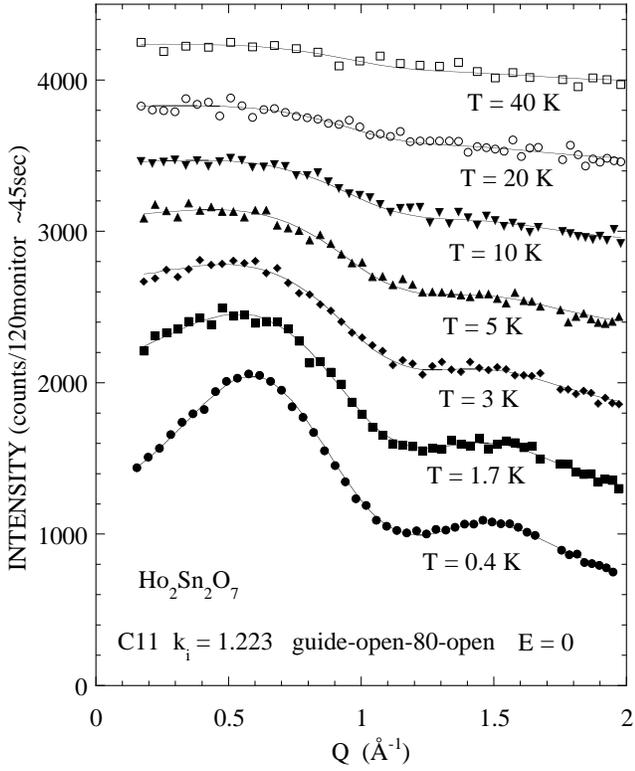}
\end{center}
\caption{
Elastic magnetic scattering of Ho$_2$Sn$_2$O$_7$ 
as a function of wave number at various temperatures.
Data of $T =$ 1.7, 3, 5, 10, 20, and 40 K are shifted by 
500, 1000, 1500, 2000, 2500, and 3000 counts, respectively 
for clarity.
Solid lines are fitted curves 
calculated using 
Eqs. (\ref{MF.wnd.suscpt}) and (\ref{CS.QEA})
of the mean field theory.
}
\label{Qscan}
\end{figure}

In the low temperature range where the system can be regarded 
as an Ising model, 
we studied magnetic short-range order, or spin correlation. 
As shown in Sec. \ref{mag.exct} the magnetic excitation 
spectra are completely elastic in the present experimental 
condition. 
In this case, double-axis data 
$\frac{d \sigma}{d \Omega} \left( {\bf Q} \right)$ 
can be measured by triple-axis $Q$-scans with $E =$ 0, 
which can be interpreted as the Fourier transform of 
the spin-pair correlation because the quasi-elastic 
approximation is applicable. 
We carried out $Q$-scans with elastic condition $E =$ 0 
using the C11 spectrometer at several temperatures 
down to $T =$ 0.4 K. 
The results of the $Q$-scans are shown in Fig.~\ref{Qscan}. 
One can see from this figure that the magnetic short-range order 
starts to develop below $T <$ 20 K. 
Although this energy scale is remarkably larger than the 
Curie-Weiss temperature 
$\theta_{\rm CW} = $ 1.8 K, 
it is comparable to the blocking energy 
$E_{\rm B} =$ 20 K. \cite{matsuhira}
By analyzing the $Q$-dependence of the intensity, which will 
be explained in the next section, we obtained 
convincing evidence that the dipolar interaction is the major
inter-spin coupling.

\begin{figure}
\begin{center}
\includegraphics[width=8.6cm,clip]{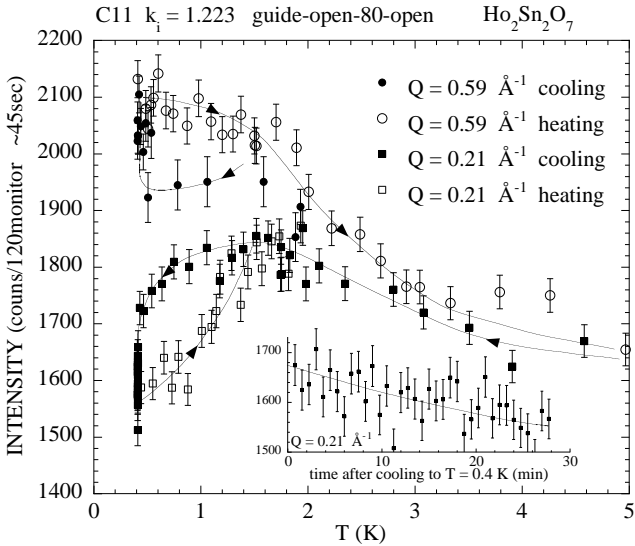}
\end{center}
\caption{
Temperature dependence of elastic magnetic scattering 
of Ho$_2$Sn$_2$O$_7$ measured at $Q =$ 0.59 and 0.21 \AA$^{-1}$
in cooling and heating conditions.
The inset shows time dependence of intensity at 
$Q =$ 0.21 \AA$^{-1}$ after temperature is cooled to 
$T =$ 0.4 K.
Solid lines are guides to the eye.
}
\label{Tdep.hysteresis}
\end{figure}

The temperature dependence of the scattering intensity 
at two typical wave numbers 
$Q =$ 0.59 and 0.21 \AA$^{-1}$ was measured in cooling 
and heating conditions, and is 
plotted in Fig.~\ref{Tdep.hysteresis}. 
We note that $Q =$ 0.21 \AA$^{-1}$ is the wave number where 
the largest variation of the intensity was observed at
low temperatures.
These data clearly demonstrate existence of thermal hysteresis 
of the magnetic scattering below $T_{\rm f} \simeq$ 1.4 K. 
This temperature is in good agreement with the onset of the
slow spin dynamics found by the AC susceptibility. \cite{matsuhira}
The thermal hysteresis synonymously implies 
time dependence of the 
intensity. 
This time dependence was really observed as shown 
in the inset of Fig.~\ref{Tdep.hysteresis}. 
The scattering intensity at $Q =$ 0.21 \AA$^{-1}$ 
varies in the time scale of the order of ten minutes, 
after the temperature is cooled to $T =$ 0.4 K. 
From these experimental facts we conclude that 
the onset of the slow spin dynamics in Ho$_2$Sn$_2$O$_7$ 
is slow formation of the short range order. 
%
\section{
Mean field analysis of magnetic elastic scattering
}
\label{MF}
%
\subsection{
Mean field theory
}
\label{MFT.subsec}
By analyzing the neutron scattering intensity 
$\frac{d \sigma}{d \Omega} \left( {\bf Q} \right)$, 
it is possible to extract information of the magnetic 
interaction parameters.
The simplest method to perform this is to utilize a 
mean field approximation of wave-number dependent 
susceptibilities. \cite{marshall,lovesey}
For simplicity of handling equations, we assume 
a spin Hamiltonian of a quadratic form of spins 
\begin{equation}
\label{hmlt.QF}
H = - \sum_{n, \nu, \alpha, n', \nu', \beta} 
J_{n, \nu, \alpha; n', \nu', \beta} 
S_{n, \nu, \alpha} S_{n', \nu', \beta}\;,
\end{equation}
where $S_{n, \nu, \alpha}$ 
represents $\alpha$-component ($\alpha$ = $x$, $y$, $z$) 
of a classical vector spin 
${\bf S}_{n, \nu} = {\bf S}_{{\bf t}_{n} + {\bf d}_{\nu}} $
($|{\bf S}_{n, \nu}|=1$) 
on a $\nu$-th site 
${\bf d}_{\nu}$ ($\nu=$ $1,2,3,4$) 
in an {\it n}-th chemical unit-cell located at 
${\bf t}_{n}$.
We further assume
\begin{eqnarray}
\label{hmlt.3T}
H &=& - D_a \sum_{n, \nu} 
\left[ 
\left( {\bf n}_{\nu} \cdot {\bf S}_{n, \nu} \right)^{2} 
- | {\bf S}_{n, \nu} |^{2} 
\right] 
\nonumber \\
&-& J_{1} \sum_{\langle n, \nu ; n', \nu' \rangle} 
{\bf S}_{n, \nu} \cdot {\bf S}_{n', \nu'} 
\nonumber \\
&+& D_{\rm dp} r_{\rm nn}^3 \sum_{\langle n, \nu ; n', \nu' \rangle}
\Bigl[
\frac{ {\bf S}_{n, \nu} \cdot {\bf S}_{n', \nu'} }
{ |{\bf r}_{n, \nu ; n', \nu'}|^3 } 
\nonumber \\
& & \; \; - 
\frac{3 
\left( {\bf S}_{n, \nu} \cdot {\bf r}_{n, \nu ; n', \nu'} \right)
\left( {\bf S}_{n', \nu'} \cdot {\bf r}_{n, \nu ; n', \nu'} \right)
}
{ | {\bf r}_{n, \nu ; n', \nu' } |^5 }
\Bigr]
\; ,
\end{eqnarray}
where
\begin{equation}
\label{def.r}
{\bf r}_{n, \nu ; n', \nu' } 
= 
{\bf t}_{n} + {\bf d}_{\nu} - {\bf t}_{n'} - {\bf d}_{\nu'}
\; .
\end{equation}
The first term is the single-ion anisotropy energy 
with the local easy-axis direction 
${\bf n}_{\nu}$ ($ |{\bf n}_{\nu}| = 1 $). 
To reproduce the strong Ising anisotropy we chose 
a large positive value $D_a$ = 810 K. 
This value is rather arbitrary and other choices of 
$D_a$ = 1620 or 405 K were confirmed not to change the 
following numerical results. 
The second term is the nearest-neighbor exchange-interaction, 
which has to be extended to further neighbors if it is required. 
The third term is the dipolar interactions of 
the Ho magnetic moments 
$\mu {\bf S}_{n, \nu}$, where $\mu$ is 
the magnitude of the moment 
$\mu \simeq$ $g \sqrt{J(J+1)} \mu_{\rm B} \simeq $ 
10 $\mu_{\rm B}$. 
The interaction constant is 
$D_{\rm dp} =$ $\mu^{2}/r_{\rm nn}^{3} \simeq$ 1.4 K, 
where 
$r_{\rm nn} =$ $a/(2 \sqrt{2})$
is the distance between two nearest-neighbor spins.
The dipolar energy 
between two nearest-neighbor spins \cite{Hertog} is 
$D_{\rm nn} =$ 
$5 D_{\rm dp} / 3 \simeq $ 2.4 K.

Mean field theories of magnetic ordering and wave-number dependent 
susceptibilities are described in standard literature for 
a Bravais lattice. \cite{marshall,lovesey}
Extending this to a non-Bravais lattice is 
a little complicated but straightforward. 
Applications to pyrochlore magnets for magnetic ordering 
are described in Refs. 
\onlinecite{gingras.cm,reimers91}, and 
\onlinecite{raju99}.
In the mean field theory, the magnetic structure 
is determined by the Fourier transform of the interaction constants
\begin{eqnarray}
\label{FT.JQ}
J_{{\bf q}; \nu, \alpha; \nu', \beta}
 =  & & \sum_{n} 
J_{n, \nu, \alpha; n', \nu', \beta} 
\nonumber \\
& & \; \times
\exp \left[- i {\bf q} \cdot 
\left( {\bf t}_{n} + {\bf d}_{\nu} - {\bf t}_{n'} - {\bf d}_{\nu'} 
\right) \right] \;,
\end{eqnarray}
where ${\bf q}$ is a wave vector in the 
first Brillouin zone, 
and eigenvalue equations 
\begin{equation}
\label{eigen.eq}
\sum_{\nu', \beta} J_{{\bf q}; \nu, \alpha; \nu', \beta}
u_{{\bf q}; \nu', \beta}^{(\rho)}
 = 
\lambda_{{\bf q}}^{(\rho)} u_{{\bf q}; \nu, \alpha}^{(\rho)}  \;.
\end{equation}
The eigenvalues 
$\lambda_{\bf q}^{(\rho)}$ ($\rho = $ $1, 2, \ldots 12$) 
and the eigenvectors 
$u_{{\bf q}; \nu, \alpha}^{(\rho)}$ 
were calculated by numerical diagonalization 
of the 12-dimensional real symmetric-matrix 
$J_{{\bf q}; \nu, \alpha; \nu', \beta}$ \cite{dp.sum} with
the normalization condition
\begin{equation}
\label{norm}
\sum_{\nu, \alpha} 
u_{{\bf q}; \nu, \alpha}^{(\rho)} 
u_{{\bf q}; \nu, \alpha}^{(\sigma) *} = \delta_{\rho,\sigma}
 \;.
\end{equation}
The system first undergoes a phase transition 
to a long-range ordered phase at a temperature
$T_{\rm C}$, 
determined by 
\begin{equation}
\label{TC}
k_{\rm B} T_{\rm C} = 
\frac{2}{3} 
[\lambda_{\bf q}^{(\rho)}]_{{\rm max}({\bf q},\rho)}
 \;,
\end{equation}
where $[\ \ ]_{{\rm max}({\bf q},\rho)}$ 
indicates a global maximum for all ${\bf q}$ and $\rho$.
The ordered magnetic structure is expressed by
\begin{equation}
\label{spin.str}
\langle S_{n, \nu, \alpha} \rangle = 
\langle S_{\bf q}^{(\rho)} \rangle 
u_{{\bf q}; \nu, \alpha}^{(\rho)} 
\exp \left[ i {\bf q} \cdot \left( {\bf t}_{n} + {\bf d}_{\nu}
\right) \right]
+ {\rm c.c.}
 \;,
\end{equation}
where $\langle S_{\bf q}^{(\rho)} \rangle$ is the amplitude 
of the modulation.

Wave-number dependent susceptibilities 
\cite{marshall,lovesey}
in the paramagnetic phase $T >$ $ T_{\rm C} $
extended for non-Bravais lattices 
are defined by
\begin{equation}
\label{wnd.suscpt}
\chi_{{\bf q}; \nu, \alpha; \nu', \beta}
=
\frac{ N^2 \mu^{2}}{ V k_{\rm B} T}
\langle S_{-{\bf q}; \nu', \beta} S_{{\bf q}; \nu, \alpha} \rangle
\; ,
\end{equation}
where $N$ and $V$ are a number of the unit cell and 
volume of the system, and
\begin{equation}
\label{Sq}
S_{{\bf q}; \nu, \alpha} 
= 
\frac{1}{N}
\sum_{n} S_{n, \nu, \alpha}
\exp \left[ - i {\bf q} \cdot \left( {\bf t}_{n} + {\bf d}_{\nu}
\right) \right]
\; .
\end{equation}
In the mean field approximation,
the wave-number dependent susceptibilities are
calculated by using the eigenvalues and 
eigenvectors
\begin{equation}
\label{MF.wnd.suscpt}
\chi_{{\bf q}; \nu, \alpha; \nu', \beta}
=
\frac{N \mu^{2}}{V} 
\sum_{\rho} 
\frac{ 
u_{{\bf q}; \nu, \alpha}^{(\rho)} 
u_{{\bf q}; \nu', \beta}^{(\rho) *}
}
{
3 k_{\rm B} T_{\rm MF} - 2 \lambda_{\bf q}^{(\rho)}
}
\;,
\end{equation}
where $T_{\rm MF}$ is the temperature defined in the 
mean field approximation.
Using this $\chi_{{\bf q}; \nu, \alpha; \nu', \beta}$
the neutron scattering cross-section 
in the quasielastic approximation \cite{marshall,lovesey} 
is expressed as
\begin{eqnarray}
\label{CS.QEA}
\frac{d \sigma}{d \Omega} 
\bigl( {\bf Q} &\;&= {\bf G} + {\bf q} \bigr)
= C
 f({\bf Q})^{2} k_{\rm B} T
\sum_{\alpha, \beta, \nu, \nu'}
\left( 
\delta_{\alpha \beta} - \hat{Q}_{\alpha} \hat{Q}_{\beta} 
\right)
\nonumber \\
&\;& \ \  \times \;
\chi_{{\bf q}; \nu, \alpha; \nu', \beta}
\cos \left[ {\bf G} \cdot 
\left( {\bf d}_{\nu} - {\bf d}_{\nu'} \right) \right]
\; ,
\end{eqnarray}
where ${\bf G}$, $C$ and $f({\bf Q})$ stand for 
a reciprocal lattice vector, a constant,
and a magnetic form factor, 
respectively. 
%
\subsection{
Application to Ho$_2$Sn$_2$O$_7$
}
\label{fitting.subsec}
In order to estimate interaction constants from 
the observed intensity data shown in Fig.~\ref{Qscan}, 
we used Eqs. (\ref{MF.wnd.suscpt}) and (\ref{CS.QEA}) 
as an experimental trial function, and performed 
least square fitting.
In this fitting, $C$ in Eq. (\ref{CS.QEA}) is treated 
as an adjustable scale factor of the intensity, 
because the sum rule of the cross-section, 
which is always satisfied experimentally,
is violated for the theoretical equations 
(\ref{MF.wnd.suscpt}) and (\ref{CS.QEA}). 
The temperature $T_{\rm MF}$ of Eq. (\ref{MF.wnd.suscpt}) 
is also one of fitting parameters, and has little meaning 
as real temperature, especially in low temperatures 
where spin correlation is not negligible. 
If the intensity data at various temperatures 
are reproduced by the same interaction constants, 
we can think that the interaction constants 
obtained by this method have real physical meaning.
We note that 
an obvious advantage of this analysis is simplicity, 
and that 
another advantage is numerical flexibility 
that any observed intensity data can be reproduced 
by introducing 
enough number of interaction constants.

\begin{figure}
\begin{center}
\includegraphics[width=8.6cm,clip]{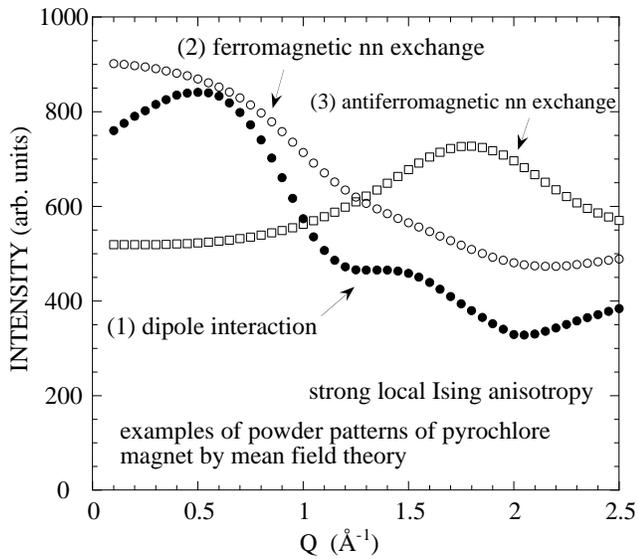}
\end{center}
\caption{
Examples of magnetic scattering calculated using 
Eqs. (\ref{MF.wnd.suscpt}) and (\ref{CS.QEA}) 
of the mean field theory.
Spin interactions of three curves 
are 
(1) dipolar interaction 
$D_{\rm dp}$ = 1.4 K, $J_{1}$ = 0;
(2) ferromagnetic nearest-neighbor exchange-interaction 
$J_{1}$ = 2 K, $D_{\rm dp}$ = 0;
(3) antiferromagnetic nearest-neighbor exchange-interaction 
$J_{1}$ = -2 K, $D_{\rm dp}$ = 0. 
}
\label{example.MF}
\end{figure}

To illustrate the theoretical scattering profile and 
to gain 
insight to what extent information can be extracted 
from the powder diffraction data, 
we calculated examples of the mean field 
$\frac{d \sigma}{d \Omega} \left( {\bf Q} \right)$
in three limiting cases: 
spins interact only via 
(1) dipolar interaction 
$D_{\rm dp} = $ 1.4 K, $J_{1}$ = 0;
(2) ferromagnetic nearest-neighbor exchange-interaction 
$J_{1} =$ 2 K, $D_{\rm dp}$ = 0;
(3) antiferromagnetic nearest-neighbor exchange-interaction 
$J_{1} =$ -2 K, $D_{\rm dp}$ = 0. 
These three examples with certain values of 
$T_{\rm MF}$ and $C$ are shown in Fig.~\ref{example.MF}. 
By comparing these curves with the observed data in 
Fig.~\ref{Qscan}, we may 
conclude that the observation excludes possibilities 
of the purely ferromagnetic or antiferromagnetic 
exchange-interaction and that the dipolar interaction 
and a small exchange interaction will account for the
observation. 

\begin{figure}
\begin{center}
\includegraphics[width=8.6cm,clip]{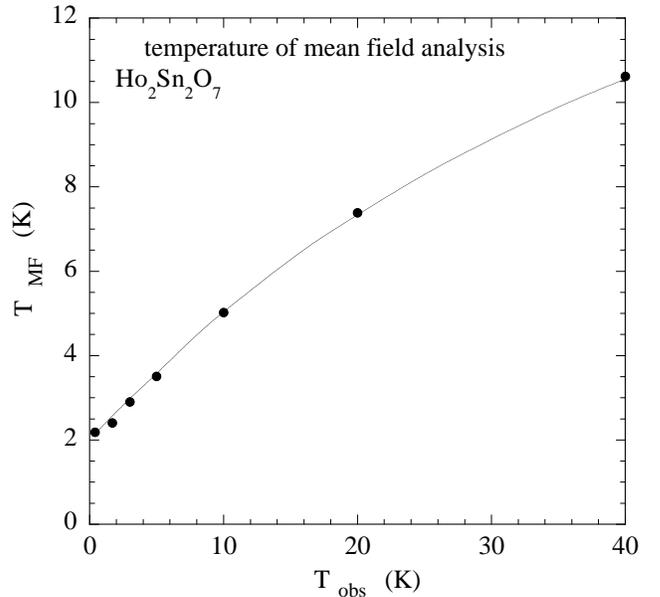}
\end{center}
\caption{
Fitted values of the temperature of the mean field theory 
$T_{\rm MF}$ in Eq. (\ref{MF.wnd.suscpt}) 
is plotted as a function of
the observed temperature $T_{\rm obs}$.
Line is guide to the eye.
}
\label{T.vs.TMF}
\end{figure}

Along this line we fitted the experimental data 
at $T = $ 0.4 K to the mean field expressions
Eqs. (\ref{MF.wnd.suscpt}) and (\ref{CS.QEA}) 
with fixed $D_{\rm dp} = $ 1.4 K and adjustable parameters of 
$J_{1}$, $T_{\rm MF}$ and the intensity scale factor. 
The fitted values are 
$J_{1}$ = 1.0 $\pm$ 0.5 K (ferromagnetic) and 
$T_{\rm MF}$ = 2.2 $\pm$ 0.1 K.
The calculated curve using the fitted parameters 
is plotted by the solid line in Fig.~\ref{Qscan}, 
which shows very good agreement. 
It should be noted that the large error of $J_{1}$
indicates that the diffraction pattern is almost
reproduced only by the dipole interaction 
and that the small discrepancy of the fitting is 
improved by adding small contribution due to $J_{1}$.
By using this value $J_{1}$ = 1.0 K, 
we tried to fit other scattering data above 
$T >$ 0.4 K with the adjustable parameters of 
$T_{\rm MF}$ and the intensity scale factor. 
The resulting fit curves are shown 
by solid lines in Fig.~\ref{Qscan}. 
The excellent fit quality ensures that the 
relative strength of the dipolar and
the exchange interactions is well established 
by the present analysis. 
In Fig.~\ref{T.vs.TMF} the fitted values of $T_{\rm MF}$ 
are shown, which seems to have little 
meaning as expected.
By following Ref. \onlinecite{Hertog}, we calculate 
the exchange energy between two nearest-neighbor spins 
$J_{\rm nn}$ = $\frac{1}{3}J_{1}$ = 0.3 $\pm$ 0.15 K. 
This is 14 \% of 
the dipolar energy $D_{\rm nn}$ = 2.4 K
of two nearest-neighbor spins. 
Therefore we conclude that the major spin-spin 
coupling in Ho$_2$Sn$_2$O$_7$ is the dipolar interaction.

%
\section{
Discussion
}
\label{disc}
As shown in the previous section, the analysis of 
the magnetic interactions using the mean field theory 
in the paramagnetic phase seems successful from a 
view point of the experimental fitting. 
However applicability of the mean field theory has to be
justified theoretically. 
In a high temperature range, Eq. (\ref{MF.wnd.suscpt}) 
can be justified as an approximation to a 
high temperature expansion. 
On the other hand, in a low temperature range of 
the order $\theta_{\rm CW} = $ 1.8 K, 
the justification is less clear. 
Equation (\ref{MF.wnd.suscpt}) may still be regarded 
as an approximation with a strongly renormalized 
temperature parameter $T_{\rm MF}$, or 
may completely lose physical meaning. 
The experimental analysis suggests correctness of 
the former possibility. 
At present, we think at least, the conclusion that 
the main spin-spin interaction is the dipolar coupling 
will not be changed, when a more precise theoretical analysis 
is made. 
It should be noted that a recent neutron scattering 
work on a single-crystal Ho$_2$Ti$_2$O$_7$ 
elucidated dipolar nature of the spin interactions 
by using a Monte Carlo simulation analysis. \cite{Bramwell.cm}
We also obtained the same conclusion 
by using the present mean field analysis
on another single-crystal neutron-data of 
Ho$_2$Ti$_2$O$_7$. \cite{kanada.nd}

\begin{figure}
\begin{center}
\includegraphics[width=8.6cm,clip]{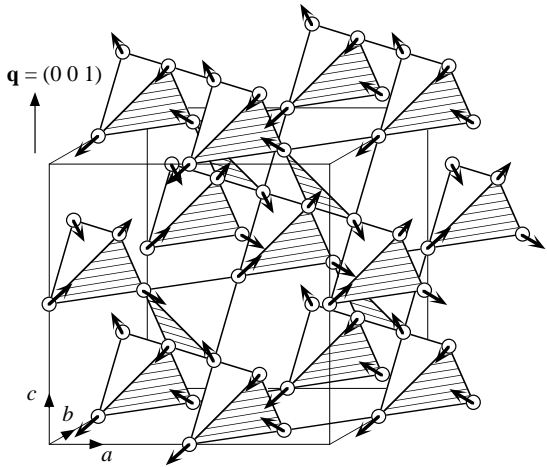}
\end{center}
\caption{
Magnetic structure with modulation 
${\bf q} = {\bf c}^{*}$ determined by 
Eqs. (\ref{TC}) and (\ref{spin.str}) of
the mean field theory. 
This is also a ground-state spin-configuration 
of the dipolar spin-ice model.
}
\label{crystal.mag}
\end{figure}

For completeness of the mean field theory and the
convenience for a reader, we would like to make a few comments 
on the long range order of Eqs. (\ref{spin.str}) and (\ref{TC}).
The maximum eigenvalue 
$[\lambda_{\bf q}^{(\rho)}]_{{\rm max}({\bf q},\rho)}$
is doubly degenerate at each X point, 
${\bf q}$ = ${\bf a}^*$, ${\bf b}^*$, or ${\bf c}^*$. 
\cite{I.MFT} 
These wave numbers correspond to the peak position 
$Q$ $\simeq$ 0.6 \AA$^{-1}$ $\simeq$ $|(001)|$
of the $Q$-scan in Fig.~\ref{Qscan}. 
Using the corresponding eigenvector 
$u_{{\bf q}; \nu, \alpha}^{(\rho)}$, 
we depict a magnetic structure with 
${\bf q}$ = ${\bf c}^*$ in Fig.~\ref{crystal.mag}, 
which is also one of the ground states. \cite{Melko}

A characteristic spin ice behavior observed in 
Ho$_2$Sn$_2$O$_7$ is the slow spin dynamics or a sort of 
spin freezing below $T_{\rm f}$ $\simeq$ 1.4 K, 
which was found by the AC 
susceptibility measurements. \cite{matsuhira} 
Quite consistent behavior is observed by the present 
work as the thermal hysteresis and the 
time dependence of the magnetic scattering. 
This implies very slow development of the short range order.
From these experimental facts, a natural question arises 
whether there exists a spin-glass phase-transition 
around $T_{\rm f}$
or around lower $T_{\rm m}$ $\simeq$ 0.75 K which was defined by
the onset of irreversibility of magnetization process, 
\cite{matsuhira}
or it is interpreted as a blocking phenomenon. 
Since the nature of $T_{\rm f}$ nor $T_{\rm m}$ 
is not well understood, the question 
is to be answered by further studies. 
We note that the thermal hysteresis and the 
time dependence of the magnetic scattering was 
observed in a spin glass system. \cite{motoya}

According to the Monte Carlo simulation study of the dipolar 
spin-ice model, \cite{Hertog} 
the spin ice behavior occurs in a wide range of 
the interaction ratio -0.8 $< J_{\rm nn}/D_{\rm nn}$. 
The isomorphic systems Dy$_2$Ti$_2$O$_7$ and Ho$_2$Ti$_2$O$_7$
have small antiferromagnetic exchange-interactions with 
$J_{\rm nn}/D_{\rm nn}$ = -0.52 and -0.22, respectively. 
\cite{Hertog,Bramwell.cm}
Ho$_2$Sn$_2$O$_7$ with 
$J_{\rm nn}/D_{\rm nn}$ = 0.14 $\pm$ 0.07
also belongs to this family. 
%
%
\section{
Conclusion
}
\label{concl}
We have investigated the frustrated pyrochlore magnet
Ho$_2$Sn$_2$O$_7$ by means of neutron scattering techniques 
using the powder sample. 
The high-resolution powder-diffraction shows that the crystal 
structure is the fully ordered pyrochlore structure with no 
detectable disorder. 
Because of the limitation of the present powder-diffraction data, 
possibilities of small amount of disorders are to be studied. 
Magnetic excitation spectra demonstrate that 
the magnetic fluctuation is almost static 
at low temperatures below $T$ $<$ 40 K, in which 
the system behaves as an Ising model. 
The crystal-field excitations observed at 
$E$ = 22 and 26 meV 
strongly suggest that the crystal-field 
state of Ho$_2$Sn$_2$O$_7$ is almost the same as that of 
the isomorphic Ising system Ho$_2$Ti$_2$O$_7$. 
By measuring elastic magnetic scattering at 
low temperatures down to $T$ = 0.4 K,
we observed only the short range order, 
which develops below $T$ $<$ 20 K. 
The wave-number dependence of the magnetic scattering 
has been analyzed successfully using the mean field theory. 
This analysis shows that spins interact mainly via the dipolar 
interaction and via the small additional 
nearest-neighbor exchange-interaction. 
Therefore we conclude that Ho$_2$Sn$_2$O$_7$ belongs to the 
dipolar-spin-ice family compounds.
The spin freezing below $T_{\rm f}$ $\simeq$ 1.4 K 
is observed as the thermal hysteresis and 
the time dependence of the magnetic scattering. 
This means unusual slow development of 
the short range order, which should be 
studied by future work. 
%
%
\acknowledgments
We would like thank 
M. Kanada, K. Motoya, M. Mekata, M. Sato, T. Sato, Y. Yasui
for valuable discussions.
%

\end{multicols}
%


\begin{references}
\bibitem{dp.sum}
In a numerical calculation, slow convergence of the
Fourier transform of the dipolar interaction has to be 
carefully treated. 
In the present work we carried out the summation of 
Eq. (\ref{FT.JQ})
in a large sphere of a radius $r =$ 10$a$ or 20$a$.
\bibitem{I.MFT}
This result is consistent with the mean field theory 
of Ref. \onlinecite{gingras.cm}, in which 
the Ising spin $\sigma_{n,\nu} =$ $ \pm 1$ is exactly treated 
in the Hamiltonian. 
In our approach the Ising spin is obtained by the limit 
$D_{\rm a}$ $\rightarrow \infty$. 
The ordered spin structure, which is not explicitly 
shown in Ref. \onlinecite{gingras.cm}, seems 
consistent with the present result.
\end{references}
\end{document}